\documentclass[a4paper,fleqn,usenatbib,useAMS]{mnras}
\usepackage{graphicx}
\usepackage[skip=2pt,font=footnotesize]{caption}
\usepackage [autostyle, english = american]{csquotes}
\usepackage{natbib}
\usepackage{mathtools}
\usepackage{capt-of}
\usepackage{tabu}
\usepackage{float}
\usepackage{hyperref}
\usepackage{color}
\usepackage[T1]{fontenc}
\usepackage{ae,aecompl}
\usepackage{newtxtext,newtxmath}
\title{ THE ACCRETION DISC-JET CONNECTION IN BLAZARS}

\author[Mukherjee et al.]{Sagnick Mukherjee,$^{1}$\thanks{E-mail: \href{sagnickm@yahoo.in}{sagnickm@yahoo.in}}
Kaustav Mitra$^{1,2}$ and Ritaban Chatterjee$^{1}$ \\
$^{1}$Department of Physics, Presidency University, 86/1 College street, Kolkata-700073, WB, India.\\
$^{2}$Department of Astronomy, 52 Hillhouse Avenue, Steinbach Hall, Yale University, New Haven, CT 06511, USA.}

\date{Last updated ; in original form }
\pubyear{2018}

\hypersetup{
        colorlinks=true,
        linktoc=all,
        linkcolor=blue,
        citecolor=blue,
        filecolor=black
}
\begin{document}
\maketitle
\begin{abstract}
The power spectral density (PSD) of the X-ray emission variability from the accretion disc-corona region of black hole X-ray binaries and active galactic nuclei has a broken power-law shape with a characteristic break time-scale ${T_B}$. If the disc and the jet are connected, the jet variability may also contain a characteristic time-scale related to that of the disc-corona. Recent observations of the blazar Mrk 421 have confirmed the broken power-law shape of the PSD of its jet X-ray variability. We model the time variability of a blazar, in which emitting particles are assumed to be accelerated by successive shock waves flowing down the jet with a varying inter-shock time-scale ($T_{IS}$). We investigate the possible relation between the characteristic time-scales in the disc and jet variability based on the above model, along with mathematically and physically simulated disc variability. We find that both the PSD of the jet and disc variability may have a broken power-law shape but the break time-scales are not related in general except only in systems with a small range of BH mass. The break in the jet and the disc PSD are connected to the interval between large amplitude outbursts in the jet ($T_{IS}$) and to the viscous time-scale in the disc, respectively. In frequency bands where multiple emission processes are involved or emission is from lower energy particles, the break in the PSD may not be prominent enough for detection.
\end{abstract}
\begin{keywords}
galaxies: active --- galaxies: jets --- accretion discs --- black hole physics --- (galaxies:) quasars: supermassive black holes --- X-rays: binaries
\end{keywords}

\section{Introduction}
Accretion on to a black hole is the primary source of electromagnetic radiation emitted by the active galactic nuclei (AGN) and the black hole X-ray binaries (BHXRBs) \citep{rees78,pringle72}. The secondary star may contribute significantly in the latter. In the AGN and BHXRBs, the presence of an accretion disc is inferred from the observations and modelling of their electromagnetic emission \citep[e.g.,][]{malkan82}. It has been speculated using the analysis of variability from these accreting systems that many features of the AGN are scaled-up version of that in the BHXRBs \citep{markowitz03,mch06}. High-resolution multi-wavelength observations of some of these accreting black hole systems have revealed the presence of a collimated outflow of magnetized plasma, termed ``jet'' \citep[e.g.,][]{rom16}. These highly energetic jets are assumed to be powered by accretion \citep{ferreira11,wu13} although the detailed mechanism of their launching and collimation is not fully understood yet \citep{dexter14,malzac04,mcnamara11}.

Consequently, a connection between the inflow in the accretion disc and the outflow in the jet may be expected. disc-jet connection has been studied in multiple BHXRBs and Seyfert galaxies \citep{miller12,fender04,fender09,soleri10,klein01,yadav06,corbel00,kanbach01}. These studies mostly use long-term multi-wavelength monitoring of these sources, revealing a connection between the decrease in brightness and spectral hardening in the X-rays with higher and/or faster flow of matter down the jet. In the broad line radio galaxies 3C 120 and 3C 111, the jet is misaligned (20$^{\circ}$ with our line of sight) \citep{jorstad05}. Hence, the optical and X-ray emission are dominated by that from the accretion disc-corona region while the radio emission is from the jet. Observations of these unique sources, have revealed similar connection between the motion of radio-bright superluminal knots seen in the pc-scale jets and the dips in the X-ray light curve \citep{marscher02,rc09,rc11,tombesi12,marscher17}. Blazars are a class of AGN with a prominent jet pointed within $\sim$10$^{\circ}$ of our line of sight. As a result, the emission from the jets is relativistically beamed and overwhelms that from the disc-corona region. Therefore, it is not possible to study a connection between the emission from the latter and the motion in the pc-scale jet. 

3C 120 and 3C 111 share another property with many BHXRBs and Seyfert galaxies. The power spectral density (PSD) of their X-ray variability has a break, i.e., the slope becomes flatter above a certain time-scale. That so-called ``break time-scale'' is consistent with the $\rm T_B$---$\rm M_{\rm BH}$---$\rm L_{\rm bol}$ relation found in a large sample of BHXRBs and Seyfert galaxies by \citet{mch06}. If the jets are indeed launched near the disc and there is a disc-jet connection as suggested by the above observations, emission variability in the jet may exhibit a similar characteristic time-scale in its power spectrum. While such a break has not been found in the power spectra of the emission variability in many blazars it has been observed in a few cases. Most recently \citet{arc18} have confirmed the presence of a break in the X-ray power spectral density of the blazar Mrk 421, which was previously hinted by other authors \citep{kat01,iso15}.  \citet{goyal18} have inferred a break time-scale in the $\gamma$-ray variability of another BL Lac object OJ 287 from their analysis of its Fermi-LAT light curve. However, it is not clear if the break in the above cases is consistent with the $\rm T_B$---$\rm M_{\rm BH}$---$\rm L_{\rm bol}$ relation above or is due to some other time-scale related to the jet emission. It is possible that the time-scale found in the jet is related to the emission mechanism(s) in the jet itself and not with the disc, or it may be related to the disc but shifted due to relativistic effects.

In this work, we develop a semi-analytical model of jet emission focusing on its variability. We examine the PSD of the simulated variability at multiple wavelengths generated by this model in order to search for the existence of any characteristic time-scale. We check whether such a time-scale may be related to certain properties or events in the accretion disc. For that purpose, we follow \citet{cowperthwaite14} to simulate the emission variability of the accretion disc. Theoretical modelling of jet emission variability with particular focus on the nature of its PSD and possible characteristic time-scale(s) present in it has been carried out most recently by \citet[][]{fin14,che16}. They find various physical time-scale in their model related to cooling, crossing, escape and acceleration of emitting particles. However, a connection of these time-scales with the physical processes in the accretion disc was not the goal of their work. \citet{malzac18} and \citet{drappeau15} have assumed that the jet variability is due to the fluctuation of the the bulk Lorentz factor of the jet plasma, which, in turn, is driven by the accretion flow. They have focused in reproducing the radio to infrared (IR) spectral energy distribution and the IR/X-ray time lag of the X-ray binary GX-339 using that model.

In {\S}2 we describe the model; we carry out the analyses, describe the results and discuss their implications in {\S}3, and in {\S}4 we present the summary and conclusions. 

\section{Model}
We model the jet as an elongated box with a circular cross-section. We divide the length of the box in multiple zones each of which fills up the cross-section of the jet, and has its own particle energy distribution and magnetic field. We assume that the emission from the electrons dominate the total jet emission and any other contribution is neglected, i.e., we are assuming a so-called ``leptonic model'' of jet emission. A shock front injects a power-law energy distribution of electrons as it passes through a zone. After the shock front passes, the electron energy distribution of each zone evolves independently of its neighbouring zones as they cool through synchrotron emission and inverse-Compton scattering (IC). We select the time interval between two consecutive shock fronts to probe various possibilities of jet variability and the disc-jet connection. We simulate the synchrotron and IC emission from the entire emission region by adding up the contribution from each zone and over an interval during which several shock fronts pass through the region. We assume the magnetic field to be the same in each zone with a value of 1 Gauss \citep[e.g.,][]{sullivan09}, and the ``seed'' photon field, which is up-scattered is from in or outside the jet. 

\subsection{Synchrotron Emission}
We average the magnetic field over all the angles that the field vector makes with the observer's line of sight. The synchrotron emission from a power-law distribution of electrons is given by \citep[e.g.,][]{rybicki86}.
\begin{equation}
j^{S}(\nu) = \int_{{\gamma}_{min}}^{{\gamma}_{max}} N(\gamma,t) d\gamma \ x \int_x^{\infty} K_{5/3}(\zeta)d\zeta 
\end{equation}
Here $j^{S}(\nu)$ is the synchrotron emission coefficient as a function of the frequency of the emitted photons. ${\gamma}_{min}$ and ${\gamma}_{max}$ are the minimum and the maximum Lorentz factor, respectively, of the electron energy distribution. $x= {\nu}/{{\nu}_c}$, where ${\nu}_c = k_1{\gamma}^2$ and $k_1~=~4.2 \times 10^6~B$, where $B$ is in Gauss. Here $K_{5/3}$ is the modified Bessel function of the second kind of order $5/3$. $N(\gamma,t)$ is the electron energy distribution in the zone. We have assumed a power-law distribution which radiatively loses its energy according to the equation:
\begin{equation}
N(\gamma,t)= N_0 \gamma^{-s} (1-{\gamma}k_{2}t)^{(s-2)}  
\end{equation}
Here, $t$ is the time after the shock has passed through that zone, $s$ is the initial power-law index of the energy distribution and $k_{2}$ is $1.3\times10^{-9}B^2$, where $B$ is in Gauss.
\begin{figure*}
\centering
\includegraphics[scale=0.05,trim={28cm 0 0 0},clip,angle=270,keepaspectratio]{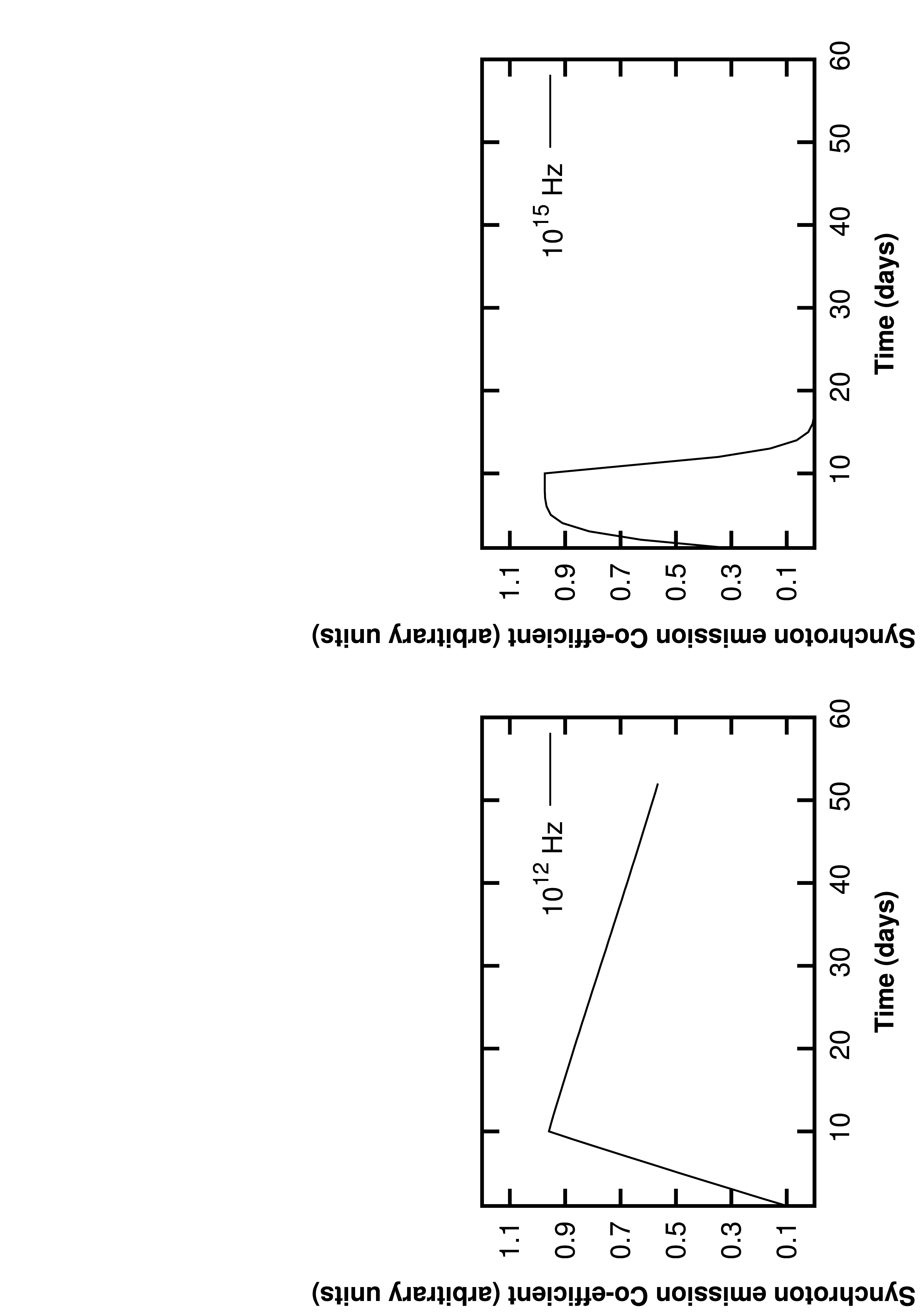}
                   
\caption{Black solid line shows the light curve at $10^{12}$ Hz (left panel) \& $10^{15}$ Hz (right panel), produced by synchrotron emission as a single shock front moves down the jet. }
\label{fig:single_zone}
\end{figure*}

\subsection{External Compton (EC) Emission}
The relativistic electrons in the jet may up-scatter ``seed" photons from outside the jet, e.g., the broad line region, to produce higher energy emission \citep{bottcher07,sikora94,coppi99,sikora00,dermer09,dermer10}. This emission mechanism is called the external Compton (EC) process. The inverse-Compton (IC) emission from the above electron energy distribution is given \citep[e.g.][]{rybicki86} by:
\begin{equation}
j^{IC}({\nu}_f) = \int_{{{\nu}_i}}  \int_{{\gamma}_{min}}^{{\gamma}_{max}} \left(\dfrac{{\nu}_f}{{\nu}_i}\right) j^{i}({\nu}_i) R {\sigma}({\nu}_i,{\nu}_f,\gamma)N({\gamma,t})d\gamma d{\nu}_i
\end{equation}
Here $j^{i}({\nu}_i)$ is the spectral intensity of the incident seed photons, ${\nu}_i$ is the frequency of the incident photons, $j^{IC}$ denotes the IC emission coefficient,  $R$ is the radius of the emission region and $\sigma$ is the scattering cross section given by:
\begin{equation}
\sigma = \dfrac{3}{32} {{\sigma}_T}\left(\dfrac{1}{{{\nu}_i}{\gamma^2}}\right)\left( 8 +2x - x^2 + 4xln\left(\dfrac{x}{4}\right)\right),
\end{equation}
where $x=\dfrac{{\nu}_f}{{{\nu}_i}{\gamma^2}}$ and ${{\sigma}_T}$ is the Thomson cross-section. We use a uniform spectral intensity distribution from $9\times10^{15}$ Hz to $1.4\times10^{16}$ Hz for the seed photons of EC emission as seen from the rest frame of the jet. 

\begin{figure*}
\centering

\includegraphics[scale=0.1,clip,angle=0,keepaspectratio]{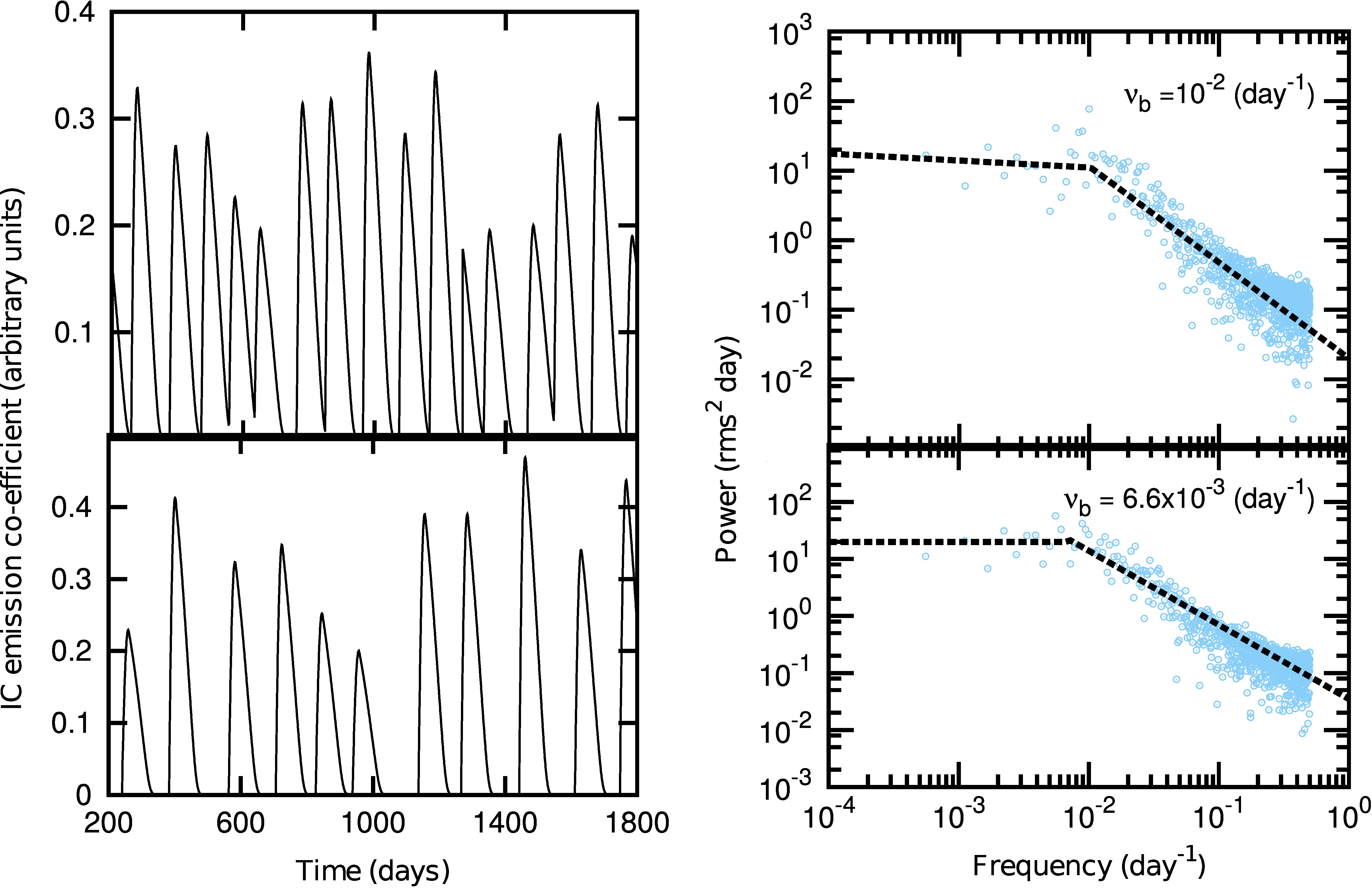}
\caption{Left Panels: Black solid lines in the upper and lower panels show the external-Compton light curve at $10^{21}$ Hz for average inter-shock time-scales of 100 and 150 days, respectively. The number of emitting particles in the jet accelerated by each shock has been drawn from a uniform random distribution of a standard deviation to mean ratio of $\sim$0.5. Right Panels: Blue open circles in the upper and lower panels denote the PSD of the light curves in the corresponding left panels. Black dashed lines show the broken power-law fit to the PSD with break frequencies of 0.01 day$^{-1}$ and $6.6 \times 10^{-3}$ day$^{-1}$, for the upper and lower panels, respectively. It is evident that the break frequency in the PSD of both the light curves matches with the inverse of their mean inter-shock time-scales.}
\label{fig:psd_100}
\end{figure*}

\subsection{Synchrotron Self Compton (SSC) Emission} 

 The Synchrotron photons produced within the jet can be up-scattered by the energized electron population of the jet itself in a process termed the synchrotron self-Compton (SSC) \citep{maraschi92,chiang02,arbeiter05}. This emission mechanism is similar to the EC emission process described in \S2.2 and given by equation 3 \& 4. $j^{i}({\nu}_i)$ for SSC emission is the spectral intensity of the synchrotron photons produced in the jet. We have included a simple treatment of the SSC process within our model, in which the synchrotron photons are upscattered within the same cell in which they are produced. While calculating the SSC emission in a given cell, the contribution of seed photons from all other cells weighted by the inverse-squared of their distance from the former is added. Hence, we find that the contribution from the other cells may be neglected and that does not  significantly affect our results regarding time variability.   

\subsection{The Accretion Disc Model}

 We follow the procedure described in \citet{cowperthwaite14} for modelling the disc variability. The disc is governed by the non-linear diffusion equation \citep{pringle81}:
\begin{equation}
\dfrac{\partial {\sigma}(R,t)}{\partial t}= \dfrac{1}{R}\dfrac{\partial}{\partial R}(\sqrt{R}\dfrac{\partial}{\partial R}(3{\nu}{\sigma}\sqrt{R}))
\end{equation} 

 The shear viscosity of the disc $\nu(R,t)$ is perturbed as $\nu(R,t)=\nu_0(1+\beta(R,t))$. The perturbation $\beta(R,t)$ has no spatial correlations which means that the perturbation at a certain radius $R_1$ is independent of that at another radius $R_2$. But $\beta(R,t)$ has a temporal correlation with a characteristic time-scale $\tau$, i.e., $\beta(R,t)$ at a fixed radius at time $t$ depends on $\beta(R,t)$ at that radius in the past back to a time $t-\tau$. The perturbation is assumed to follow an Ornstein-Uhlenbeck (OU) process which ensures that the perturbations are stochastic in nature and all the correlation properties stated above are met. $\beta(R,t)$ is assumed to follow the differential equation \citep{cowperthwaite14}:  
\begin{equation}
d\beta(R,t) = -\omega(R)\beta(R,t)dt + \sqrt{\dfrac{\omega(R)}{\omega_0}}dW(R,t)
\end{equation}
Here, $\omega(R)$ is the correlation frequency, which is assumed to be the local viscous frequency ($\omega(R)={\nu_0}/R$) of the disc and the equation is normalized by the viscous frequency of the innermost disc ($\omega_0$). $dW(R,t)$ is a Gaussian noise. The perturbation in the outer parts of the disc are correlated for a much longer time than the inner regions of the disc.

We divide the disc into 100 annuli with each annulus having a thickness of 0.1 code units. We solve the disc equation numerically with stochastic perturbations in viscosity which follow an OU process. We use a time-resolution of 0.2 time units. We apply all the considerations described by \citet{cowperthwaite14} in order to ensure numerical stability of the solution.

\section{Results}
\subsection{Nature of the Characteristic time-scales in the Model Jet Variability}
The multi-wavelength light curves generated by the above model are shown in Figure \ref{fig:single_zone}. As time passes, more and more electrons are energized by the shock front and the emission increases. Photons of higher frequencies are produced by higher energy electrons which lose energy faster. Therefore, emission at different frequencies is distributed over various spatial extents behind the shock front. Only a small region behind the shock front contains high-energy electrons emitting high-frequency radiation. This yields a flat-top light curve as shown in Figure \ref{fig:single_zone} right panel. On the other hand, photons which are created by lower energy electrons take a much longer time to decay after the shock has passed the emission region. Therefore, emission from the lower energy electrons increases continuously till the shock reaches the end of the region and then decays slowly, as shown in Figure \ref{fig:single_zone} left panel. The nature of the light curve depends on the geometric parameters such as the size of the emission region, or a gradient in the number density of the electrons or that in the direction-averaged magnetic field magnitude along the axis of the jet. The flow of multiple shocks down the jet one after another gives rise to the long-term light curves of the jet. The flares due to multiple shock fronts become more distinct, if observed at higher frequencies as the cooling time-scale is shorter for high-frequency bands and the flux decays quickly from its peak soon after the shock front passes through. 
\begin{figure*}
\centering
\includegraphics[scale=0.05,clip,angle=0,keepaspectratio]{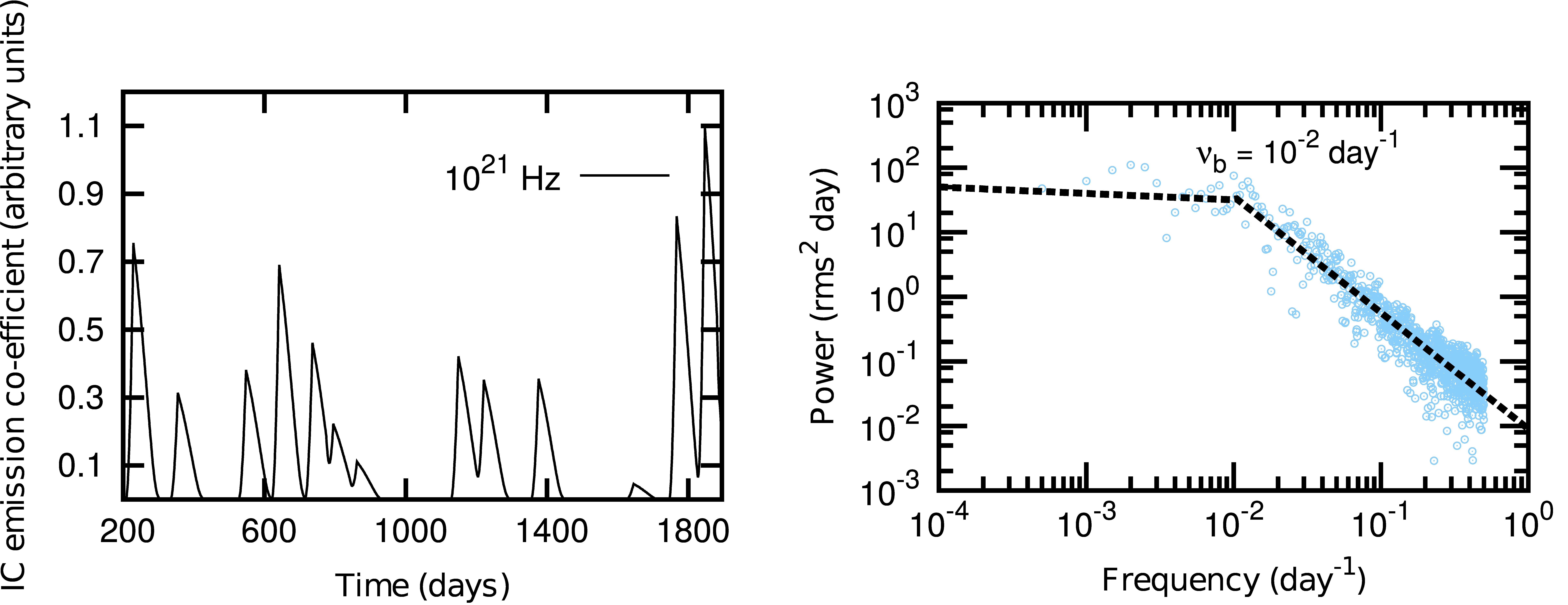}

\caption{Left Panel: Black solid line exhibits the light curve at $10^{21}$ Hz photon frequency, produced by EC emission where the dips in the 3C 120 X-ray light curves are assumed to be the shock launching times and the number of emitting particles in the jet accelerated by each shock is correlated to the amplitude of the dip. Right Panel: Blue open circles denote the PSD of the light curve shown in the left panel. Black dashed line shows the broken power-law fit to the PSD. }

\label{fig:psdjetobs}
\end{figure*}   

\begin{figure*}
\centering
\includegraphics[scale=0.05,clip,angle=0,keepaspectratio]{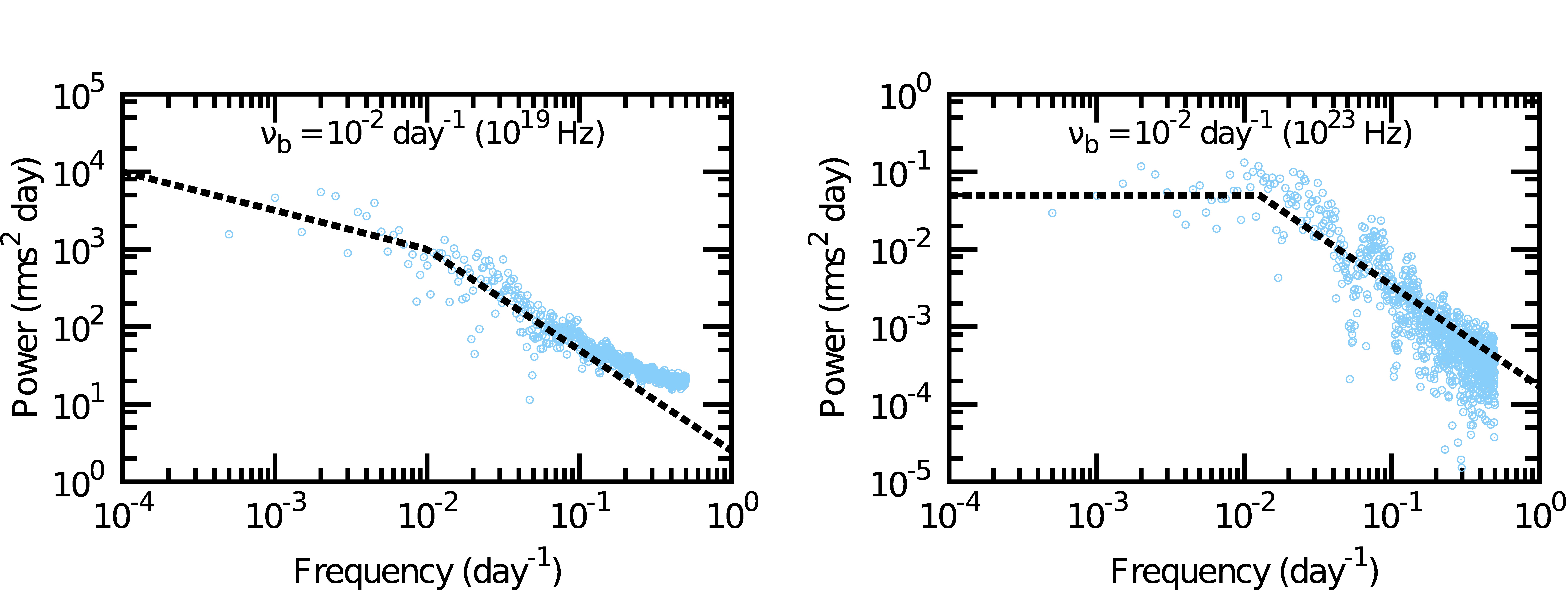}

\caption{Blue open circles denote the PSD of light curves at $10^{19}$ and $10^{23}$ Hz in the left and right panel, respectively. The break time-scale for both are the same but the difference in the slope is not prominent for the lower energy photons (left panel) as compared to the higher energy photons (right panel). Black dashed lines show the broken power-law fits to the PSD.}

\label{fig:sharp_psd}
\end{figure*}

We generate two sets of light curves in which the mean time interval between the passing of consecutive shock fronts is different (100 days \& 150 days) and the number of emitting particles in the jet accelerated by each shock has been drawn from a uniform random distribution of a standard deviation to mean ratio of $\sim$0.5 for both the cases. We calculate the power spectral density (PSD) of these light curves. Those light curves along with their PSD are shown in  Figure \ref{fig:psd_100}. In both the cases, it is evident that there is a break in the PSD, i.e., the slope steepens below a certain time-scale. We fit the PSD with a broken power-law model, in which the index changes at a certain frequency. The ``break frequency'' corresponds to a break time-scale ($T_B$) which is the inverse of the temporal frequency. We find that  $T_B$ is equal to the respective mean inter-shock time-scale in both the cases. This implies that the break in the PSD of the light curves is mainly caused by the repetitive shocks going down the jet and the interval between the passing of these shock-fronts.

In order to study the above connection in more details, we vary the set of times when a shock front reaches the emission region in our model. In the radio galaxies 3C 120 and 3C 111, a new radio knot is observed to move down the jet at superluminal speed a few weeks following a dip in the X-ray emission \citep{rc09,rc11}. We use the time of such dips in the X-ray light curves of 3C 120 between 2002 and 2007 as the new set of times for launching of the shock front in our model. The number of particles accelerated by each shock front is correlated with the amplitude of the dip. The time between the dips varies significantly but the mean interval is $\sim$3 months. The light curve and the resultant PSD are shown in Figure \ref{fig:psdjetobs}, where we find that the shape of the PSD is similar to what we find above and the break time-scale is approximately the mean time interval between the X-ray dips.

There are three important time-scales in our model: the acceleration time-scale $\tau_{ac}$, the cooling time-scale $\tau_{cool}$ and the inter-shock time-scale $\tau_{IS}$. $\tau_{ac}$ is the time the shock takes to travel down the length of the emission region. It is the same for all the frequencies of emission and is independent of the processes of emission. $\tau_{cool}$ depends on the emission frequency. Emission produced by lower energy electrons has a higher $\tau_{cool}$. $\tau_{IS}$ is the mean time between consecutive shocks, which may be related to the accretion disc activity of the system. The PSD captures the amount of variability at each of these time-scales. The higher frequency regime of the PSD is dominated by the variability at time-scales comparable to $\tau_{cool}$ while the lower frequency end of the PSD exhibits the variability at the time-scales comparable to $\tau_{IS}$. The break in the PSD may be due to the different variability processes at these two time-scales. One of the several characteristic time-scales found by \citet{che16} in their model PSD, the so called ``acceleration decay'' time-scale ($\tau_{\rm decay}$), may have a similar physical origin to $\tau_{IS}$. On the other hand, for lower energy emission, $\tau_{IS}$ and $\tau_{cool}$ are comparable and the PSD fails to capture the difference between the two different types of variability and hence the break in the power spectrum becomes less sharp. This has been demonstrated in Figure \ref{fig:sharp_psd}. The right panel in Figure \ref{fig:sharp_psd} shows the PSD at 10$^{23}$ Hz. The slopes before and after the break frequency are $-0.09$ and $-1.3$ respectively. The lower frequency (below the break frequency) slope becomes $-0.5$ while the higher frequency slope remains the same for the left panel showing the PSD at 10$^{19}$ Hz, which makes the break difficult to identify.
\begin{figure*}
\centering
\includegraphics[scale=0.036,clip,angle=0,keepaspectratio]{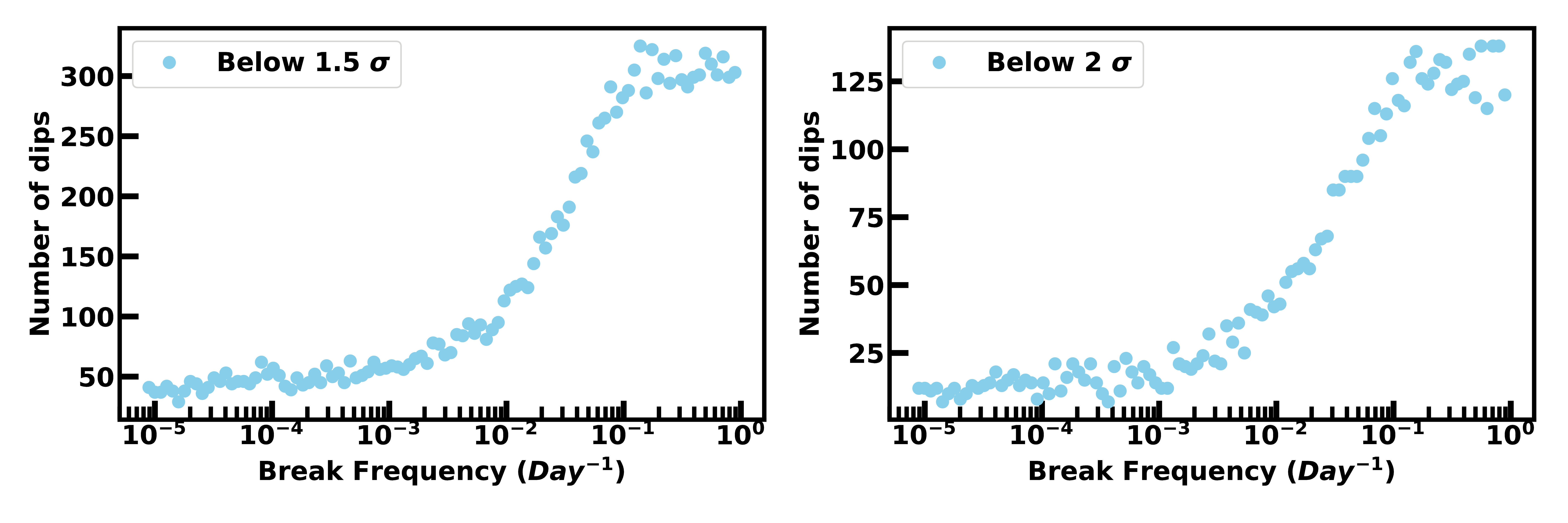}

\caption{Left Panel: Blue solid circles show the number of dips in the disc light curve \textit{vs} the break time-scale in the disc light curve simulated using the algorithm by \citet{timmer95}. Dips are defined as times when the flux is 1.5$\sigma$ below the moving average (defined in {\S}3.2) of the light curve. Right Panel: Blue points denote the same, where dips are defined as times when the flux is 2$\sigma$ below the moving average of the light curve.}
\label{fig:discvsbreak15}
\end{figure*}  

While broken power-law shaped PSD has been found in the X-ray variability of many BHXRBs and Seyfert galaxies it has been unambiguously detected only in one blazar, namely, Mrk 421 \citep{kat01,iso15,arc18}. To detect such a break a well-sampled broad-band PSD is necessary, which in turn needs light curves containing variability at hours to years time-scale. Such data sets may not be available for many blazars. The break time-scale of the jet PSD is found to be invariant across all emission wavelengths. However, there may be some other reasons for the non-detection of such breaks. The wavelength of emission at which a broken PSD can be detected depends on the SED of the jet. We calculate the PSD of variability from 10$^{10}$ Hz to 10$^{24}$ Hz emission frequency. The PSD of the emission at frequencies between 10$^{10}$ and 10$^{11}$ Hz follows a single power law as these are mainly synchrotron emission from low-energy electrons. Moderate to high- energy synchrotron emission and low-energy SSC emission contribute to the emission at frequencies from 10$^{12}$ to 10$^{15}$ Hz. The PSD is broken in this frequency window. From 10$^{16}$ to 10$^{18}$ Hz, the PSD is again a single power law. This is low-energy EC and high-energy SSC emission. For all frequencies greater than 10$^{18}$ Hz, the PSD is broken and the main emission mechanism is EC from moderate to very high-energy electrons. It is clear that the nature of the PSD depends on the energy of the electron population contributing at the frequency of emission. Furthermore, if emission at a certain wave band, e.g., 0.2$-$10 keV X-rays or 0.2$-$300 GeV $\gamma$-rays, is contributed by multiple radiation mechanisms or particles with a large range of energies, or if the outbursts at the relevant time-scales are overlapping then the break in the PSD may be blurred and may not be detected. Frequencies at which both high-energy synchrotron and lower energy SSC or high-energy SSC and lower energy EC contribute to the emission significantly, the nature of the PSD depends on the relative spectral intensities of each process. The spectral intensity of the EC emission depends on the source of seed photons for the EC emission mechanism. The detection of a break in the GeV PSD and non-detection of a similar break in the PSD of the variability at other wave bands in the case of the blazar OJ 287 \citep{goyal18} may be due to the above reason. We analyze the weekly binned 0.1-300 GeV light curve of OJ 287 during 2008 to 2018 as provided by the LAT team through their website \footnote{\url{https://fermi.gsfc.nasa.gov/ssc/data/access/lat/msl_lc/}}. We divide the light curve into 12 equal segments and calculate the local mean and standard deviation ({$\sigma$}) of each of these segments. We study two definitions of flares as points which are 1.0 $\sigma$ or 0.9 $\sigma$ above the local mean. We find the average time-scale between outbursts to be 175 and 167 days for the 1.0 $\sigma$ and 0.9 $\sigma$ definitions, respectively. We choose the fraction of standard deviations carefully such that most of the outbursts are identified by these choices. Lower values of the fraction identify very small fluctuations as outbursts and higher values miss significant outbursts and identify only the very large flares.

In addition, the jets may have additional source of variability at the short time-scale such as turbulent magnetic field and density. This may introduce additional power in the variability at the high-frequency part of the PSD, which can cause the break to be less prominent.  The light curves are usually not regularly sampled and contain large gaps in them due to various constraints in observations and that makes the detection of such breaks even more difficult.         

\subsection{Possible Connection between Characteristic time-scales in the Model Disc and Jet Variability}
\subsubsection{Jet Variability is Driven by the Dips in the Disc Light Curve} 

We find that the jet light curve in our model has a break and the break time-scale is given by the mean time interval between the passing of shock fronts through the emission region. The passage of shock fronts may follow dips in the disc light curve as observed in 3C 120 and 3C 111 and some BHXRBs. This is a connection between phenomena happening in the disc and that in the jet. To study this further using our model, we test whether there is a relation between the nature of the disc variability and the mean interval between the dips. For this purpose, we simulate disc light curves using the algorithm prescribed by \citet{timmer95}. In this algorithm, we select the break frequency, and the power-law indices below and above it, and generate a light curve, the PSD of which has the selected properties. We define the ``dips'' in such a light curve as values that are 2 standard deviation ($\sigma$) below a moving average (mean of a local segment where each local segment has a width of 500 time units and an overlap of 100 time units with the previous and the next segment), and local minimum of each of the dips is defined as the time when a shock front passes through the emission region in our model. In this manner, given a simulated light curve we can generate a set of times when shock fronts will be generated. This is crucial for the test as the break time-scale is essentially the mean time interval between consecutive shock fronts.

 We select $-1.0$ and $-2.5$ as the power-law indices below and above the break frequency ($\nu_B$), which are the standard values \citep[e.g.,][]{rc09,rc11,remillard06}. We vary the value of $\nu_B$ of the PSD of the simulated light curve from $10^{-5}$ day$^{-1}$ to 1 day$^{-1}$ and repeat the same analysis in order to find various sets of times for the passage of shock. We smooth the light curves with a Gaussian kernel using a smoothing length of 10 days in order to avoid defining very sharp fluctuations as dips because it might not be realistic that a very short-time-scale dip consisting of only one or two data points below the defined level may signify a shock event down the jet. We plot the number of dips in the TK95 light curves \textit{versus} the break frequency of the PSD of the simulated light curves in Figure \ref{fig:discvsbreak15} (left and right panels for the case when dips are defined as 1.5$\sigma$ and 2$\sigma$ below a moving average, respectively). The number of dips remains constant with $\nu_B$ for lower values of break frequencies, increases with $\nu_B$ above a certain value of  break frequency and then saturates to a high value for higher $\nu_B$. The nature of the PSD for very small break frequency is very similar to a very steep power law and hence one expects to find slowly varying light curves with long dips and hence the number of dips/shock events are lower. The PSD is a very flat power law when the break frequency is very high and the light curves are rapidly fluctuating giving rise to a larger number of dips/shock events. For intermediate values of $\nu_B$, we see a gradual increase in the number of dips/shock events as we transition from very steep to very flat power-law PSD with increasing $\nu_B$. Changing the break frequency of the TK95 light curve is motivated by the relation of the disc break frequency with M$_{BH}$. We have established that the jet break time-scale is related to the inter-shock time-scale of the jet which we directly connect to the inter-dip time-scale of the disc. As the number of dips, and the inter-dip time-scale, does not vary with the break frequency of the disc PSD at very low and very high values of $\nu_B$, we do not expect any correlation between the disc and the jet break timscales for higher and lower $M_{BH}$. For intermediate $M_{BH}$ this correlation may be present, i.e., the disc and the jet break time-scales may be related for this range of BH mass. 

We vary the post break-frequency power-law slope of the PSD and check for any relation between that and the number of dip/shock events for various break frequencies. We do not find any significant correlation between the two.

\begin{figure*}
\centering
\includegraphics[scale=0.1,clip,keepaspectratio]{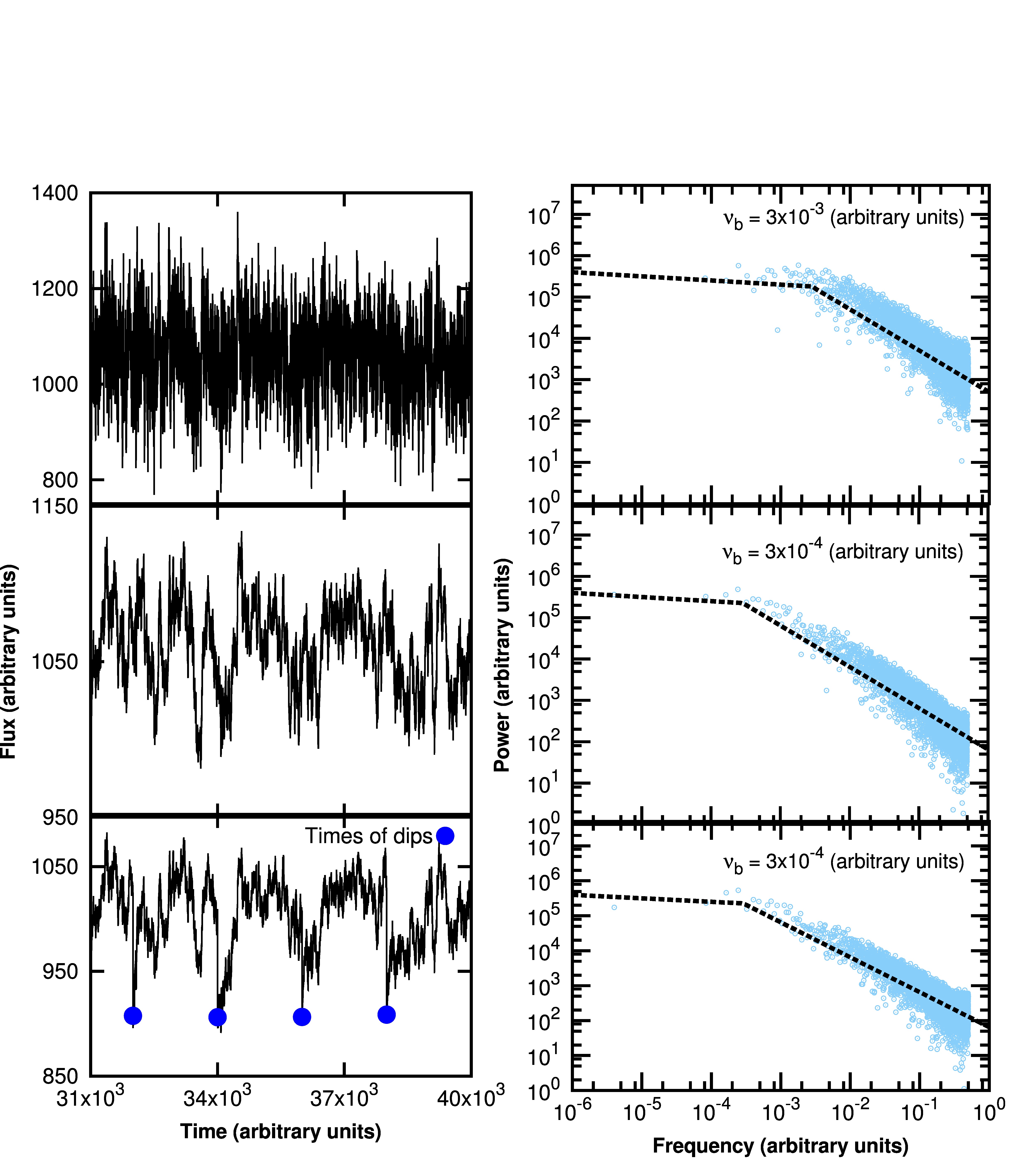}

\caption{Left Panels: Black solid lines at the top and middle panel show the disc light curve produced assuming the viscous frequency of the inner disc to be 10$^{-3}$ and 10$^{-4}$ (arbitrary units), respectively. That at the bottom panel shows the same as the middle panel with the addition of dips to the light curve by suppressing the accretion rate in the inner parts of the disc. The blue solid circles denote the times at which the light curve dips. Right Panels: Blue open circles denote the PSD of the disc variability shown in the corresponding left panel. Black dashed lines show the broken power-law fit to the PSD with a break near a frequency of 3x10$^{-3}$, 3x10$^{-4}$, 3x10$^{-4}$, respectively at the top, middle, and bottom panels. It is evident that the break frequency does not change due to the addition of the dips.}
\label{fig:psdnormal}
\end{figure*}

\subsubsection{Disc Variability is Simulated using a Physical Model}
In order to further search for any connection between the PSD break time-scale in the disc and jet variability, we study the stochastic variability of the disc discussed in \S2.4. We simulate the disc light curves using the model with different viscous frequencies and calculate the PSD of those light curves. It is evident from Figure \ref{fig:psdnormal} that the PSD follows a broken power law and the break frequency is related to the viscous time-scale at the inner parts of the disc, similar to that found by \citet{cowperthwaite14}.

In order to simulate the prominent dips as observed in the disc-corona light curves of 3C 120 and 3C 111, we suppress the surface density in the inner disc by 20\% at times drawn from a Gaussian distribution with a mean of 2000 time units and a standard deviation of 100 time units. This mean interval is different from the viscous time-scale of the disc. The light curve simulated from the above model and its PSD are shown in Figure \ref{fig:psdnormal}. The shape of the PSD of this simulated light curve with the prominent dips still follows a broken power law and the break frequency remains the same as that of the model disc variability without the dips. This implies that the PSD break frequency for the disc variability is not related to the mean time interval between successive dips but to the viscous time-scale of the disc. This indicates that if the disc variability is caused by a process similar to what has been proposed by \citet{cowperthwaite14} and the jet variability is related to the dips in the disc emission, while both may have a broken power-law shaped PSD the characteristic break time-scale of these two processes may not be the same or have the same physical origin. A connection between the phenomena in the disc and the jet does not necessarily cause a similar characteristic time-scale in the variability.

\section{Summary and Conclusions}
 We have developed a semi-analytical model of the jet to study the time-variability of its emission. The time variability arises from the acceleration of particles due to the passing of a shock-front through various zones of the emission region in the jet and subsequent cooling of the particles through synchrotron and inverse-Compton processes. Multiple shock-fronts flow down the jet one after another, with varying time intervals. In addition, we simulate emission variability of the accretion disc for further probing the connection between the two. \\

1. The PSD of the model light curves follows a broken power law with a characteristic break frequency, which is found to be directly linked to the mean time interval between the consecutive shocks.  \\

2. In the radio galaxies 3C 120 and 3C 111, dips in the X-ray emission from the disc-corona region have been observed to be followed by the motion of superluminal radio knots down the jets \citep{rc09,rc11}. We simulate the jet variability by sending shocks down the model jet at the times of the X-ray dips observed in 3C 120. The PSD of the resultant jet emission variability has a broken power law shape as above, in which the break frequency corresponds to the mean time interval between the dips. However, this break time-scale ($\sim$ 100 days) is much longer than the break time-scale ($\sim$ 1.29 days) obtained from the X-ray PSD of 3C 120 by \citet{rc09}. \\

3. We simulate disc-corona light curves with a PSD having a broken power law shape using the algorithm by \citet{timmer95}. We define the significantly low states of the light curve as dips. We vary the break frequency ($\nu_B$) of these light curves across five decades and find that the number of dips, or equivalently shocks down the jet, is uncorrelated to $\nu_B$ if the latter is very low or very high. But the number of dips/shock events increases with $\nu_B$ for a small range of intermediate values of the latter. As the break time-scale of disc PSD is correlated with $M_{BH}$, this analysis suggests that we may not find correlations between the jet and the disc break time-scales for higher or lower $M_{BH}$ but may expect a correlation for a small range of intermediate $M_{BH}$ values. \\

4. We generate disc variability following the physical model of \citet{cowperthwaite14}. The PSD of the variability has a broken power law shape with a break frequency consistent with the viscous time-scale in the inner disc. We simulate prominent dips in the above light curve by suppressing the accretion rate in the inner disc. The PSD of these modified light curves has a broken power law shape with a break time-scale identical to that of the variability without the dips. This implies that the PSD break frequency for the disc variability is not related to the mean time interval between successive dips but to the viscous time-scale of the disc. \\

5. If the jet variability is observed at a wave band, in which there is significant contribution from multiple emission mechanisms or radiation emitted by lower energy particles such that the consecutive outbursts significantly overlap, the break in the PSD may be blurred and may not be detected with irregularly sampled light curves as are usually available. \\

We finally conclude that while a connection between phenomena in the disc and that in the jet have been observed in some AGN and BHXRBs, and PSD of the variability of both the disc and the jet may have a broken power law shape, break time-scales in these two cases are not the same or related in general except an indirect relation in a small range of BH mass as described in item (3) above. The break time-scale in the disc variability may be related to the viscous or other time-scales in the disc while that in the jet may be related to the mean time interval between large amplitude outbursts in its emission. This implies that the break in a PSD detected in a jet-dominated source, such as a blazar, may be related to the motion of shock waves down the jet that causes large amplitude outbursts or some other physical process that determines the time-scale below which the amplitude of variability significantly decreases. Furthermore, the detection of a break in the jet PSD at a certain wave band implies that the emission in that band may be dominated by a single mechanism and by the high-energy part of the particle distribution, which may be tested by studying its spectral energy distribution. This is consistent with finding such a break in the X-ray variability in the blazar Mrk 421, in which the X-rays are dominated by synchrotron radiation by very high-energy electrons in the jet \citep[e.g.,][]{abd11}.

\section{Acknowledgements}
SM acknowledges the DST INSPIRE fellowship and JBNSTS fellowship. KM received support from the KVPY fellowship and JBNSTS fellowship. RC thanks Presidency University for support under the Faculty Research and Professional Development (FRPDF) Grant. RC thanks IUCAA for their hospitality and usage of their facilities during his stay in 2018 summer as part of the university associateship program. RC received support from the UGC start-up grant. The authors thank the anonymous referee for several suggestions, which helped in improving the draft.\\
 
\bibliographystyle{mnras}
\bibliography{manuscript}{}
\end{document}